\documentclass{sig-alternate-05-2015}

\usepackage[utf8]{inputenc}

\usepackage{amsmath}
\usepackage{hyperref}
\usepackage{graphicx}
\usepackage[]{array}
\usepackage[]{bm}

\usepackage[]{booktabs}
\usepackage{bookmark}
\usepackage[final]{listings}
\usepackage{lscape}
\usepackage{mathtools}
\usepackage{paralist}
\usepackage{flushend}
\usepackage[]{float}

\newcommand{\specialcell}[2][c]{%
  \begin{tabular}[#1]{@{}c@{}}#2\end{tabular}}

\begin{document}

\title{Mobile Encryption Gateway (MEG) for Email Encryption}

\numberofauthors{4}
\author{
\alignauthor
Gregory Rehm\\
    \affaddr{University of California, Davis}\\
    \email{grehm@ucdavis.edu}
\alignauthor
Michael Thompson\\
    \affaddr{Argonne National Laboratory}\\
    \email{thompsonm@anl.gov}
\alignauthor
Brad Busenius\\
    \affaddr{Argonne National Laboratory}\\
    \email{bbusenius@anl.gov}
\and
\alignauthor
Jennifer Fowler\\
    \affaddr{Argonne National Laboratory}\\
    \email{jfowler@anl.gov}
}
\date {}
\maketitle

\begin{abstract}
\par Email cryptography applications often suffer from major problems that prevent their widespread implementation. MEG, or the Mobile Encryption Gateway aims to fix the issues associated with email encryption by ensuring that encryption is easy to perform while still maintaining data security. MEG performs automatic decryption and encryption of all emails using PGP. Users do not need to understand the internal workings of the encryption process to use the application. MEG is meant to be email-client-agnostic, enabling users to employ virtually any email service to send messages. Encryption actions are performed on the user's mobile device, which means their keys and data remain personal. MEG can also tackle network effect problems by inviting non-users to join. Most importantly, MEG uses end-to-end encryption, which ensures that all aspects of the encrypted information remains private. As a result, we are hopeful that MEG will finally solve the problem of practical email encryption.
\end{abstract}

\section{Introduction}
\par Many previous attempts to make cryptographically secure email applications user friendly and widely available have failed. For varied reasons, these applications have not reached widespread adoption. The most common issue associated with email encryption lies in the difficulty that novice users face when trying to effectively encrypt their email, and even with special training, most average users are not able to effectively encrypt their emails \cite{whitten1999johnny}. Other detractions from the mass usability of email encryption include the inability to verify the identity of contacts, a lack of end-to-end encryption, and network effect issues.
\par To solve these problems we propose the Mobile Encryption Gateway (MEG); a secure, free, and highly usable alternative to all other previous email encryption technologies. First, we discuss why previous attempts to encrypt email en-masse have failed. We then describe how MEG will address all these deficiencies, including how MEG's architecture allows for a secure, end-to-end encrypted system that is email-client-agnostic, enabling users to keep their preferred email provider by installing an additional plugin for their browser or email application. We will then show MEG's performance does not constitute a barrier to usability. Finally, we discuss MEG's usability characteristics and how it enables users to perform all the necessary actions to securely encrypt messages without forcing users to understand the encryption process.

\section{Background}
\par A majority of the communications that traverse the Internet today are encrypted (HTTPS, iMessage, WhatsApp, etc.). Encryption helps ensure that our bank records and store purchases are protected. The basis for one of the most common types of encryption technology on the Internet dates back to 1976 when Whitfield Diffie and Martin Hellman proposed public key cryptography. Since then encryption has slowly become ubiquitous in electronic communications. However, after the 2013 Snowden revelations the world was shown just how vulnerable their data is to malicious actors. Since then, there has been a flurry of activity towards encrypting all communications over the Internet \cite{wired-encrypted-traffic}; and yet three years later, despite the rise in usage of Transport Layer Security (TLS) by companies like Google \cite{gmail-tls-report}, most email traffic remains unencrypted end-to-end. The reasons for this are complex, but often boil down to a lack of usability of email encryption services and network effects. The average user does not have the time, money, willingness, or technical knowledge to fully encrypt their email especially when free and intuitive services like Gmail and Yahoo email offer hassle free plain text communications \cite{garfinkel2005make}. These issues, in turn, create a feedback loop where network effects kick in. If one party encrypts their email but the recipient cannot decrypt the email, then  others are less likely to encrypt their communications; since if they did their contacts would not be able to read them \cite{dingledine2006anonymity}.

\subsection{What is Required?}
Throughout the years dozens of applications have been developed in the attempt to enable widespread email encryption. To ensure end-to-end encryption of communications these applications must do the following four things:

\begin{itemize}
  \item \textit{Sign} the message with the sender's private key. This validates the identity of the sender
  \item \textit{Encrypt} the message to ensure only the desired recipients can read it.
  \item \textit{Verify} the original signature from the sender.
  \item \textit{Decrypt} the message when received by the recipient so the message can be read.
\end{itemize}

Certain encryption methods perform some of these tasks differently. Signing in PGP works where anyone with a PGP key can sign anyone else's key. This validates identity through a concept called \textit{web of trust}. In contrast, X.509 certificates are only signed by a Certificate Authority (CA), and as a result, are trusted based on reputation of the signing CA.

\subsection{S/MIME}
\par The reigning standard for email encryption is Secure/Multipurpose Internet Mail Extensions (S/MIME). S/MIME has the ability to perform end-to-end encryption for email and is backed by CAs that can validate the identity of the receiver. The problem with S/MIME is the issue of certificate creation. Creation of X.509 certificates is a centralized process imposed by the CAs \cite{garfinkel2005johnny}. Another downside is CAs charge expense for their services, which is anathema to many users accustomed to free services like Gmail and Outlook. Even if a user is willing to go through the cost and effort of obtaining a certificate, network effects must be taken into account. If an email recipient does not have an S/MIME certificate themselves, then email cannot be encrypted. To alleviate this problem, there exist email services such as Ciphermail that have built-in CAs to automate S/MIME usability issues. However, payment is required for these services to offset cost of obtaining X.509 certificates \cite{ciphermail-gateway}. Also, third-party paid services do not always make their code auditable, ensuring experts cannot validate their system is end-to-end encrypted. Corporations often provide the service of supplying S/MIME certificates for their employees, but this only ensures encryption will occur between members of the same corporation, and not necessarily with parties outside corporate boundaries. As a result, average users cannot rely on corporations or paid services to fix S/MIME's deficiencies in accessibility.
\par Another approach, Key Continuity Management (KCM), created by Simson Garfinkel, promises to provide automated generation of self-signing S/MIME certificates \cite{garfinkel2005johnny}. Theoretically, this would mean any user could generate an S/MIME certificate instantaneously and free of charge. In this manner, KCM could resolve network effects over time. The problem with KCM is many email providers do not accept self-signed certificates because they have an inherent security vulnerability; there is no third party verification of identity. To fix this, Garfinkel argued that users can determine whether to trust where the message is coming from, similar to the way SSH works \cite{garfinkel2005make}. However, this approach does not constitute a feasible solution because the vast majority of users do not understand the purpose of self-signed certificates and therefore do not understand the security risks of using them \cite{downs2006decision}. When study participants were prompted with security decisions in a study on HTTPS (deciding whether or not to accept dubious certificates) users frequently chose to accept them \cite{callegati2009man}. Downs findings imply that malicious users could impersonate a company the user does business with to get them to accept a self-signed certificate \cite{downs2006decision}. For this reason, major email clients do not even accept the use of self-signed certificates \cite{force-use-of-self}. Furthermore, KCM could place users at greater security risk for attacks like phishing, given that they trade built-in spam security filtering for encrypted email.
\subsection{PGP}
%\textit{web of trust}
The main alternative to S/MIME is called Pretty Good Privacy (PGP). In contrast to generating X.509 certificates, creating PGP keys is a decentralized process and can be performed for free in a matter of seconds. PGP is able to validate identity through a concept named \textit{web of trust}. In practical terms, \textit{web of trust} works through key signing; if a trusted friend signs the key of another person that they (presumably) trust, in the \textit{web of trust} you could logically trust the identity of the person your friend trusts \cite{zimmermann1995official}. Assuming that a user has properly curated their \textit{web of trust}, PGP is just as capable as S/MIME validating the identity of participants in a communication \cite{furnell2013usable}.
\par Despite the advantages of being decentralized and free, PGP is notoriously difficult to use. In a famous 1999 study named \textit{Why Johnny Can't Encrypt}, researchers found that most study participants could not even encrypt their emails using graphical user interface (GUI) tools for simplifying PGP actions. The few that were able to encrypt their communications were prone to displaying their private key in plaintext; defeating the entire purpose of encryption \cite{whitten1999johnny}. A follow-up study in 2006 with improved GUI tools had no more success with study participants than the original trial \cite{sheng2006johnny}. As a result it is likely that users will continue to experience difficulties when attempting to manually encrypt their email.
\par Paid services like Hushmail are available to automate PGP, but they too suffer from lack of end-to-end encryption and have unauditable code \cite{ciphermail-gateway,hushmail,eff-scorecard}. Most importantly, these services charge money, meaning widespread adoption may not occur.
\par Open source email encryption applications using PGP are also available. One such example is named OpenKeychain. It is superb in terms of usability and accessibility, but only Android email clients are supported and OpenKeychain does not allow use of clients that the user has grown accustomed to \cite{openkeychain}. This is just one such example, and many other free PGP applications like Enigmail follow the same route of requiring usage of a specific email client \cite{enigmail-handbook}.
\begin{figure*}[h]
    \includegraphics[scale=0.50]{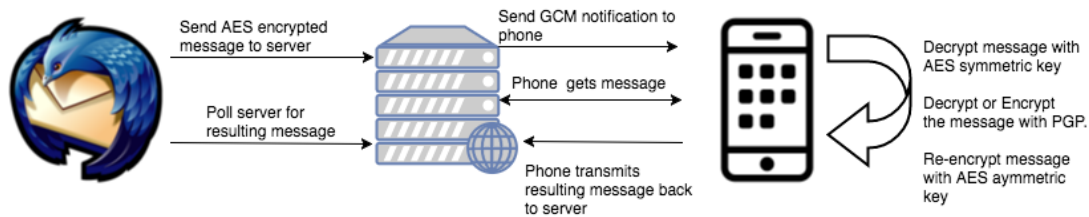}
    \centering
    \caption{MEG: Message Flow}
    \label{fig:Flow}
\end{figure*}

\subsection{MEG}
\par The goal of MEG is to address all the problems associated with using PGP, S/MIME, and provider-based encryption schemes. First and foremost MEG will use PGP because it is free, decentralized, and can validate contact identity via a \textit{web of trust}. MEG also aims to alleviate the usability troubles surrounding PGP by automating the processes of key generation, signing keys, the encryption, and decryption of messages for the user. These solutions will enable users to enjoy the security encryption provides without worrying about the low-level details of how it works; thereby eliminating the problems encountered in \textit{Why Johnny Can't Encrypt} \cite{whitten1999johnny}.
\par MEG will ensure user data remains private. End-to-end encryption will be built into MEG as necessary. Meanwhile, the user's private key will stay on their phone. Only the phone's owner will have the ability to decrypt messages. The actual encryption of emails will be provided by the MEG Android app on a mobile device which will serve as the email gateway. MEG will relay the email over a secure channel back to the user's email client, which can forward the email to its intended destination.
\par MEG will automatically handle the process of signing emails when they are sent with the user's private key. When an email is received MEG will verify the message signature using the sender's public key. If an email does not have a valid signature it will not be decrypted.
\par Finally, to combat the network effects within encryption, MEG will provide email-based invitations to anyone without the service. The process will be similar to how social networks like Facebook grew in popularity, through electronic word of mouth \cite{trusov2009effects}.

\section{Architecture}
\par The architecture of the MEG system is what enables it to be secure, private, and flexible enough to ensure users can still use their existing email clients. The entire system consists of three components:
\begin{enumerate}
    \item A mobile app that performs all PGP related actions (e.g. key generation, encryption, decryption, and signing).
    \item A email client plugin that sends messages to the phone for encryption or decryption before they are either sent to a recipient or read.
    \item A server component that acts as a PGP public keystore, a database for revocation keys, and a message broker between the client and android device
\end{enumerate}
At the time this paper was prepared, a Thunderbird email plugin and an Android app were under development for MEG, with an iPhone app and Gmail plugin planned in the future.
\par Upon installing the Android app users are directed to the installation page, where they input requisite information to generate both a PGP public/private key pair and a revocation key. The revocation key is meant to be used if a user loses their phone or if their password compromised. Activating this key can ensure that no one else will trust the user's public/private key pair in the future.
\begin{figure}[H]
    \centering
    \includegraphics[scale=.42]{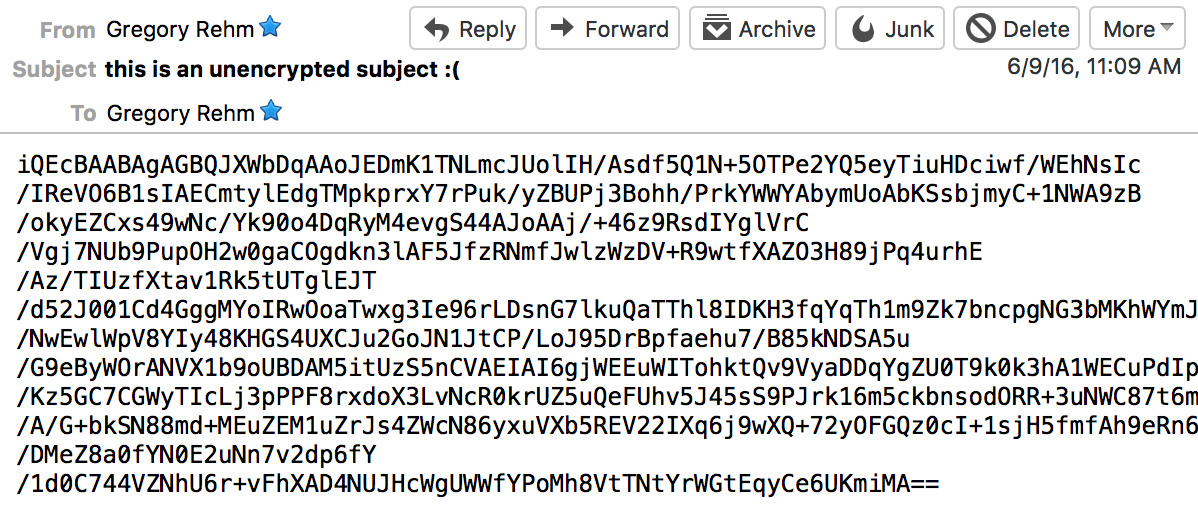}
    \caption{An Encrypted Message in the email Client}
    \label{fig:Encrypted-client}
\end{figure}
\par To ensure that communications between the mobile app and the email client plugin remain secret, we ensure that all communications between these two components are encrypted via an AES-256 symmetric key. Because of the AES encryption, even if an attacker gained access to the MEG server component, the user's communications would remain encrypted. AES encryption ensures that MEG is encrypted in transit (i.e. data in motion). PGP encryption ensures email is encrypted at rest. Combining these two encryption techniques ensures that MEG employs end-to-end encryption.

\par Because mobile devices are frequently networked behind routing devices or ISP-controlled networks that block incoming communications, the server component will assist in routing messages. The mobile app is engineered to send requests to the server for messages awaiting processing. To accomplish this, we first send a Google Cloud Messaging (GCM) notification to the phone from the server to alert it that a message is awaiting processing. Upon receiving this notification, the phone will retrieve the message and perform whatever PGP action it needs to. After processing completes, the phone will send the message back to the server for the client app to retrieve. The complete flow of how a message is transported from the client plugin is illustrated in Figure \ref{fig:Flow}.

\par Since all encryption actions are performed on the phone, email data remains private. The user's private key is also stored on the phone. The private key is already protected by a password, and the proliferation of data encryption on smart-phones ensures that the user's private key remains doubly secure. In case the phone is lost, or if the PGP key password is compromised, the user can revoke a key by requesting the server to revoke the key. This will cause an email to be sent by MEG to the user's email account for validating that they actually made the request. This step ensures that there is a guard against malicious actors revoking MEG PGP keys.

\begin{figure}[H]
    \hspace*{-0.5cm}
    \centering
    \includegraphics[scale=.45]{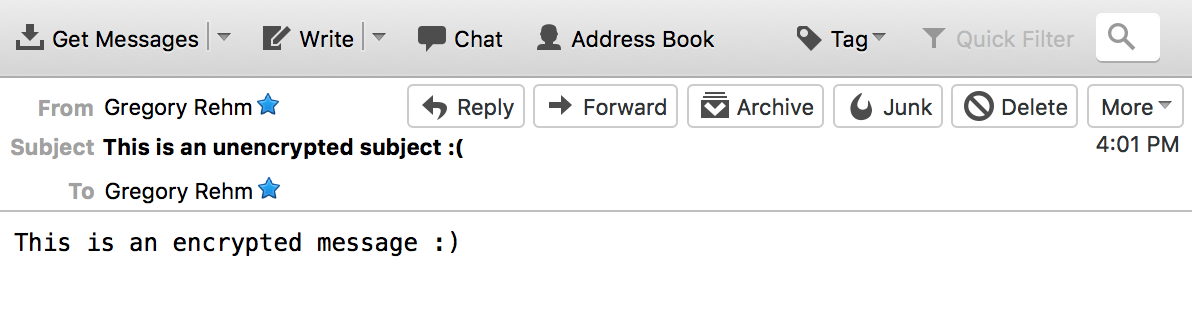}
    \caption{The Same Message, but Decrypted after Javascript Injection}
    \label{fig:Decrypted-client}
\end{figure}
\par The email client plugin is engineered to guarantee that MEG stores no plaintext data for malicious actors to view. Emails are stored on Google/Yahoo/Microsoft/etc. servers in encrypted form, ensuring the email provider does not have access to its contents. In order to decrypt a message, the MEG plugin looks for a special header attached to all MEG emails and then transmits those messages to the phone for decryption. When the messages arrive back on the client, the email is decrypted with the AES-256 symmetric key and then displayed on screen through Javascript injection. This display is ephemeral and only lasts as long as the user has the message window open. Figure \ref{fig:Encrypted-client} shows how the user would initially view their email at rest in the client. Figure \ref{fig:Decrypted-client} shows that exact same message after the Javascript injection occurs.

\section{Usability}
Our primary aim was to make MEG as usable as possible to ensure that people with little technical training would be able to encrypt their email. In the MEG system, there are two components that users are meant to interact with: the email client plugin and the mobile app. First, we discuss how the mobile app's minimal user interface (UI) makes performing PGP actions as easy as possible for users. We also show that the generation of symmetric key information does not require any technical skill. Finally, we discuss how the email client plugin enables people to use the same email services they are familiar with to send encrypted email and that this plugin requires small amounts of effort to utilize.
\subsection{Mobile App Usability}
\par
\begin{figure}[H]
    \centering
    \includegraphics[height=8cm,width=5cm]{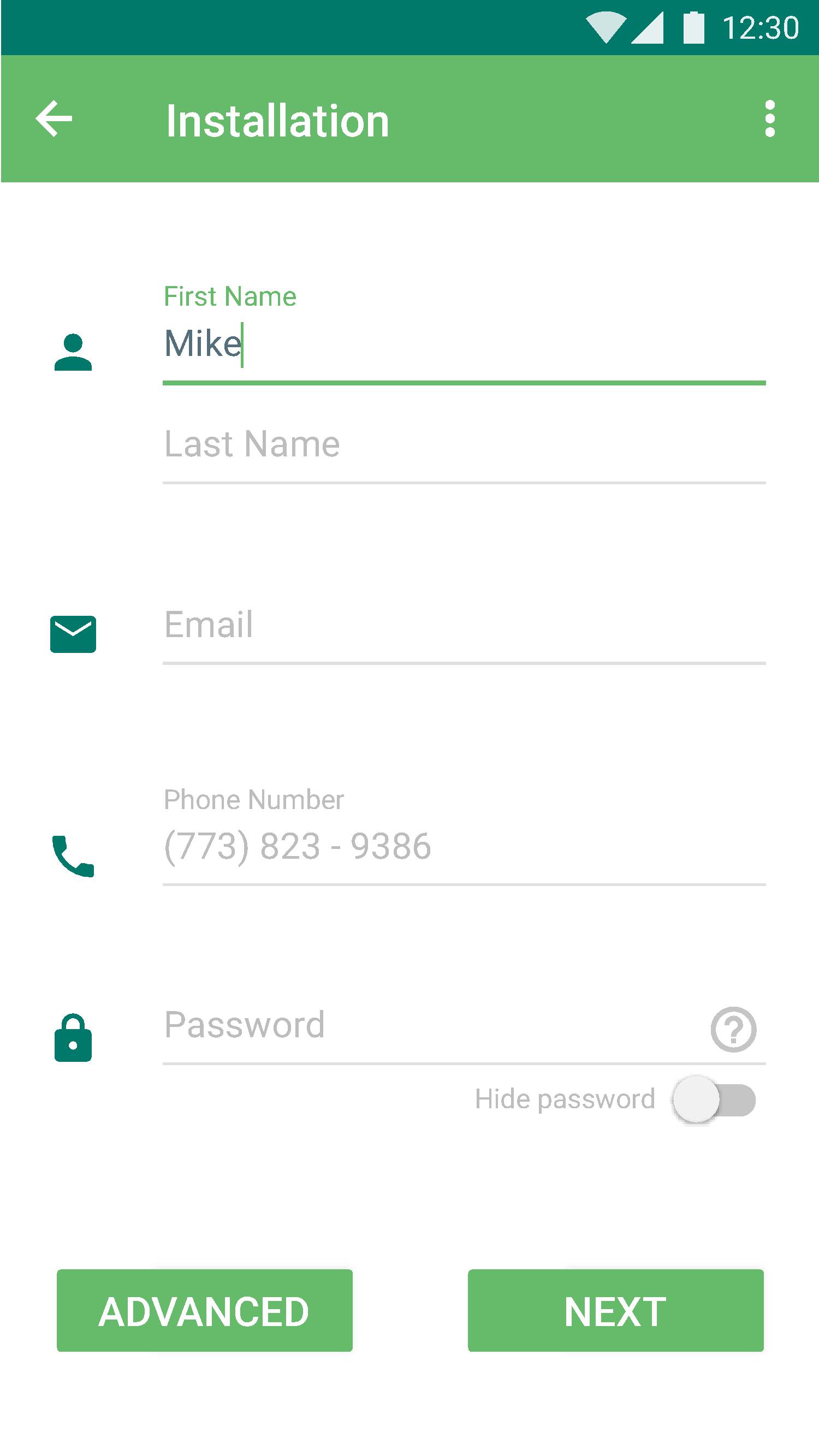}
    \label{fig:installation}
    \caption{MEG's Installation Page}
\end{figure}

 The MEG mobile app is where all PGP encryption, decryption, signing, and key generation actions take place. Fortunately, users do not need to perform any manual work to accomplish actions like encrypting data, and as mentioned earlier, MEG handles all this for us automatically. To decrease confusion when using the app, only four different UI panels are navigable. The app is designed to make generating a PGP key as simple as possible. Figure \ref{fig:installation} shows that only five pieces of information are needed from the user to generate a PGP public/private key pair: first and last name, phone number, email, and password. After a user inputs this information, a PGP public/private key pair is generated.
\par
\begin{figure}[H]
    \centering
    \includegraphics[width=8cm,height=6cm]{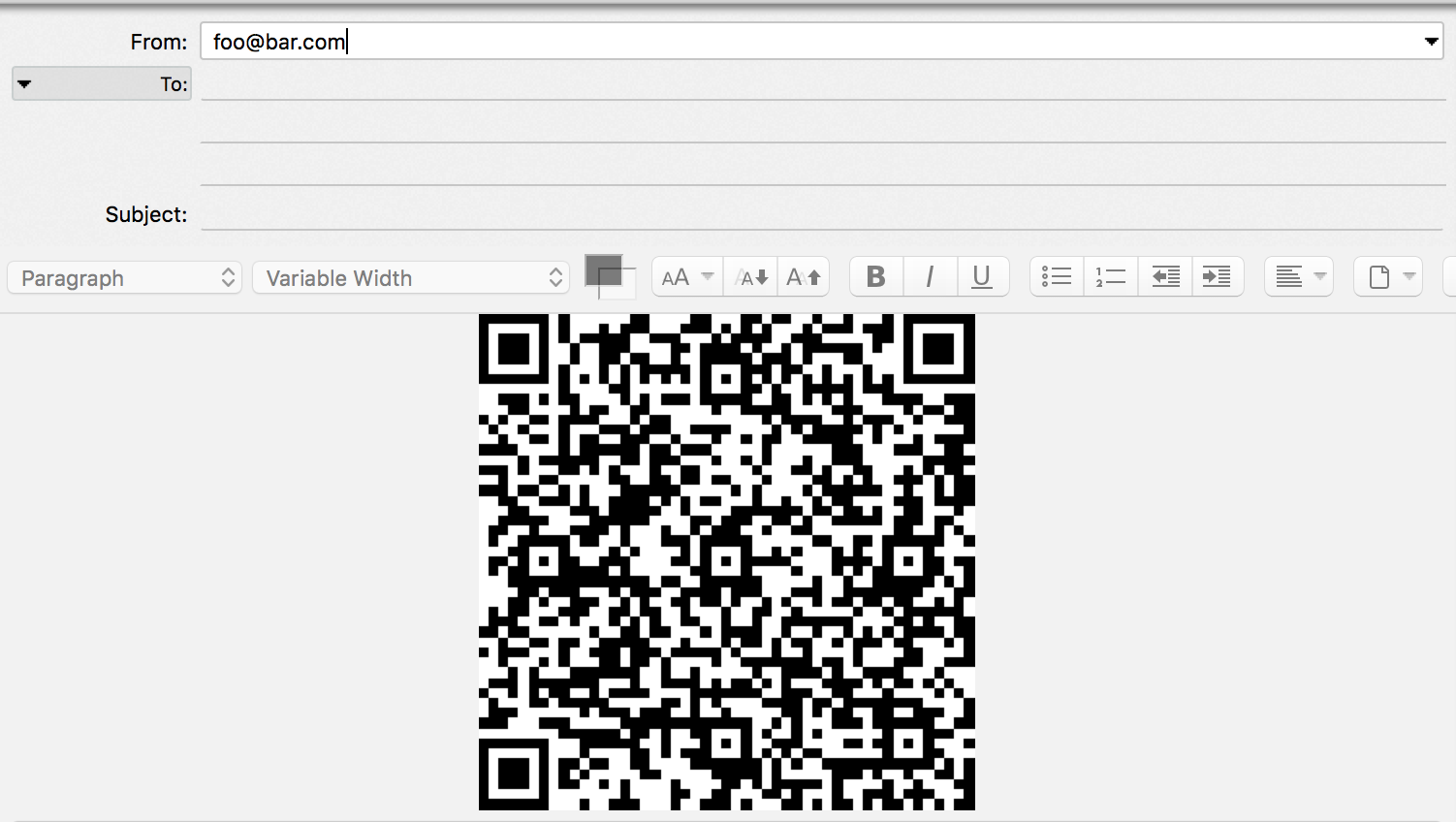}
    \caption{A QR Code Ready to be Scanned}
    \label{fig:qr}
\end{figure}

 After generating a public/private key pair the mobile app needs to scan a QR code to authenticate an email client plugin. In security-based applications, QR codes have even been used for physical access control and they can securely transmit PGP information \cite{qrcode-authentication,qrcode-key-distribution}. The reason this mechanism is secure is because a QR code does not need to be transmitted over a potentially insecure medium like the Internet, and instead, can stay securely with the user. In MEG, we leverage this feature and use a QR code to transmit AES-256 symmetric key information from the email client plugin to the mobile app. This symmetric key information is then used to encrypt communications between the email plugin and the the phone. The symmetric key is generated in the email client plugin as shown in Figure \ref{fig:qr}. After the QR code is scanned the image is removed from the screen ensuring that malicious actors cannot gain access to the symmetric key information. The main advantages of using a QR code to transmit symmetric key information are as follows: (1) it is both secure and easy to use, (2) most users are well acquainted with using the cameras on their phones; (3) no additional password input in the email client is required for decrypting AES-256 encrypted messages returning from the phone. This ensures that the user only needs to remember one password when using MEG, the password for the PGP private key on their phone.
\par One negative aspect of some email encryption applications is that users must continually input their password each time they want to send and receive email. This is especially true of server-based applications because caching a decrypted private key would mean that the application could access the user's encrypted files and the service is not end-to-end encrypted. In MEG, because all encryption actions occur on the user's phone, and the phone is in physical possession of a user, MEG can avoid the user having to re-enter a password each time an email is encrypted or decrypted. As a result, MEG can implement  in-memory private key caching on the mobile app. Upon starting the MEG mobile app, users will need to enter their password to unlock their private key. The key is cached and can be reused as long as the app stays open, ensuring that users can encrypt and decrypt their email by having their phone on and MEG open. If the mobile app is closed by the user or the operating system (OS) for some reason then the user will need to log back into the mobile app. Ultimately, this will be detected if a user attempts to either send or decrypt an email on the client. The client will then alert the user that they need to log back into the mobile app to complete their action.
\subsection{Client Usability}
\par While the mobile app performs all PGP actions, the client plugin is where the majority of user interaction will occur in MEG. MEG was designed so that users would have to perform as few actions as possible to encrypt their messages.

\begin{figure}[H]
    \centering
    \includegraphics[scale=.4]{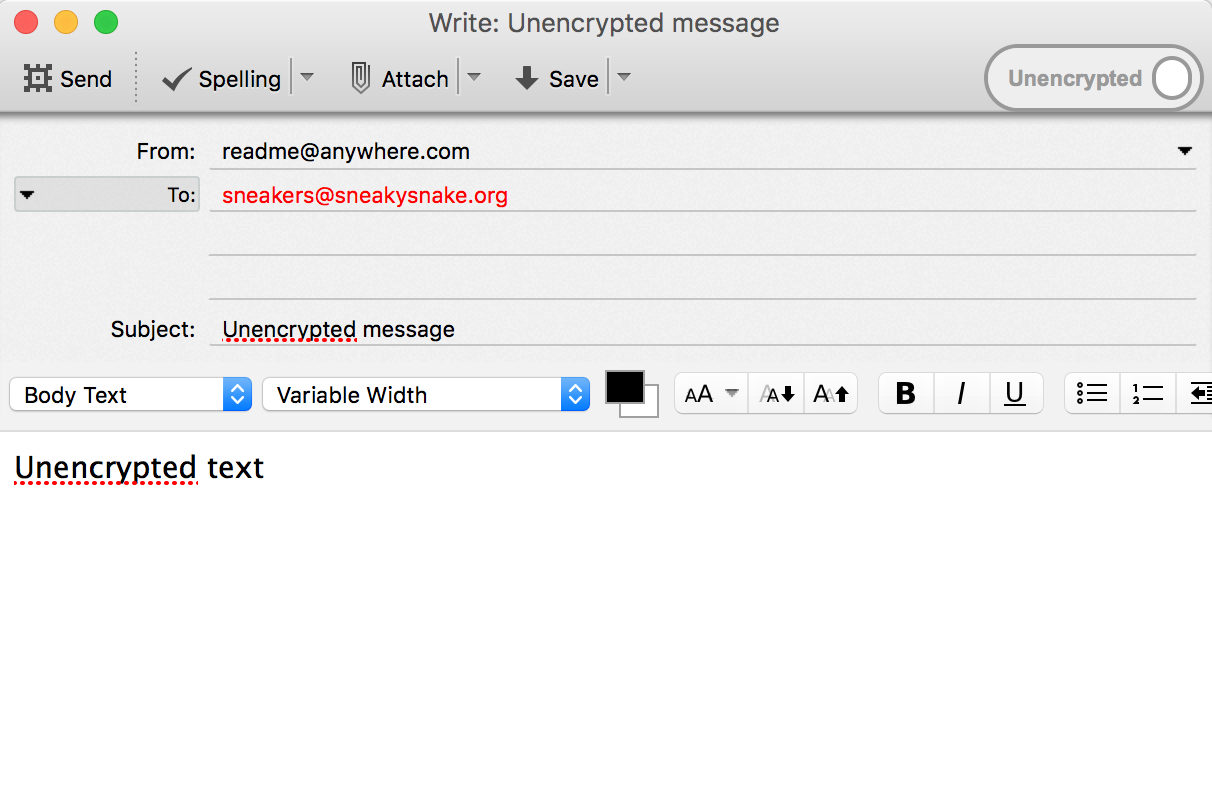}
    \caption{Client UI for Sending Unencrypted Messages}
    \label{fig:unencrypted-ui}
\end{figure}

\par The two main actions a user needs to perform on the email client are (1) generating a QR code so symmetric key data can be sent to the phone, and (2) sending encrypted email. The QR code is generated on the first attempt the user makes to send a piece of encrypted email. Once scanned, the QR code is removed from screen and user can begin to send encrypted email.
\par
\begin{figure}[H]
    \centering
    \includegraphics[scale=.4]{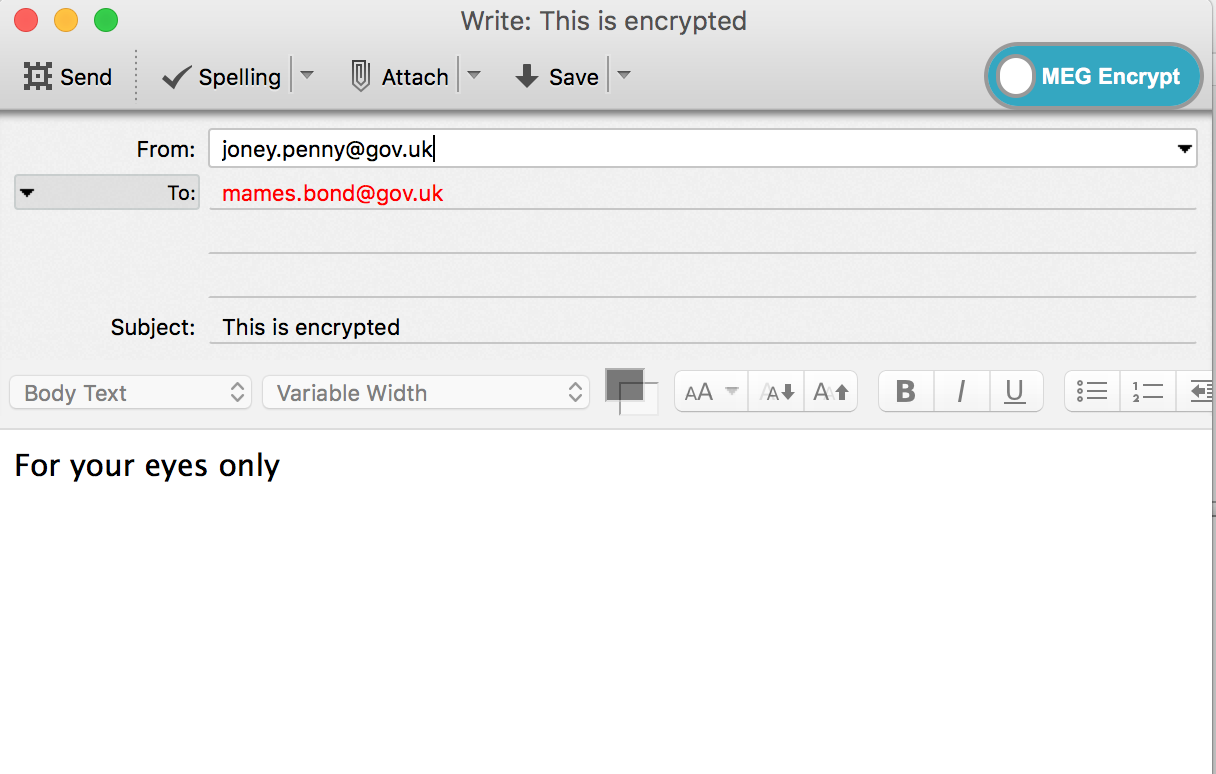}
    \caption{Client UI for Sending Encrypted Messages}
    \label{fig:encrypted-ui}
\end{figure}

 Encrypted email is toggled through a check-box at the top of the email client. Users are always given the option to leave their email unencrypted in case their recipient is unable to read the encrypted email or in the case that their mobile device lacks network connectivity. Figure \ref{fig:unencrypted-ui} shows the implementation of the basic Thunderbird UI where a user can continue to send unencrypted mail with MEG. Figure \ref{fig:encrypted-ui} shows how encrypting an email can be performed by toggling the encrypt checkbox.
\par Since existing mechanisms are used for sending email in the client, learning a new routine for sending an encrypted email is not necessary. This is a hurdle that the default encryption mechanism on Thunderbird and some email plugins such as Enigmail introduce \cite{enigmail-handbook}. Conversely, MEG only requires users to toggle a button for their email to be encrypted.
\par To tackle network effects associated with encryption, if User A sends an encrypted email to User B without MEG, then User A will send an email asking User B to sign up for MEG services. As stated earlier, this is similar to how social networks like Facebook and Twitter became so popular, through electronic word of mouth \cite{trusov2009effects}. This approach will help alleviate the network effect problems associated with encrypted email.

\section{Performance}
\par MEG utilizes three separate components, in serial fashion, to encrypt and decrypt messages. So it would be reasonable to assume that performance of this system could be burdensome to a user. However, after thorough benchmarking of the client, server, and mobile application, we believe MEG imposes no significant speed impediment to the user.
\par Benchmarking is renown as a difficult process to perform effectively \cite{gregg2013systems}. One of the key issues is that different machines may perform differently in benchmarking. So, based off previously used methodology \cite{altun2006clustering}, in Table \ref{tab:envs} we show the environment we ran our benchmarks.
\begin{table}[h]
    \centering
    \begin{tabular}{c c c}
        \toprule
        \textbf{Component} & \textbf{Architecture} & \specialcell{\textbf{Operating}\\\textbf{System}} \\ \toprule
        \textit{Mobile} & \specialcell{4 core Krait 400 CPU\\2.3GHz 2G RAM} & Android 6.0.1 \\ \hline
        \textit{Server} & \specialcell{2 core Intel i5 CPU\\2.67GHz 8G RAM} & \specialcell{Ubuntu 14.04.5\\LTS} \\ \hline
        \textit{Client} & \specialcell{2 core Intel i5 CPU\\2.90GHz 8G RAM} & OSX 10.11.5 \\
        \bottomrule
    \end{tabular}
    \caption{Benchmarking environment}
    \label{tab:envs}
\end{table}
 A preset email of 300 characters was used in each trial. This email did not vary from trial to trial. It is also important to note the client, mobile, and server components were located in the same city at the time of testing. The testing we performed was for the end to end process of encryption and decryption of emails. In all trials our timing is recorded in seconds. Since Mozilla Thunderbird lacks automated performance testing tools we had to perform our testing in an auto-manual method that we will discuss in the component testing subsections. Unfortunately, this also means it was only feasible to conduct 20 trials each for both encryption and decryption tasks.
\subsection{Mobile Benchmarking}
In MEG, mobile is the easiest component to test since the actions performed, encryption and decryption of messages, are easily isolated in the application's code. We tested these processes end to end, from the time the phone receives a message to the time the phone returns its transformed product back to the server. Table \ref{tab:mobile} shows the number of trials ($n$), the median, average ($\mu$), and standard deviation ($\sigma$) of the times recorded in the trials for both decryption and encryption of emails.
\begin{table}[h]
    \centering
        \begin{tabular}{c | c c c c}
        \toprule
        & \textbf{$n$} & \textbf{median} (s) & \textbf{$\mu$} (s) & \textbf{$\sigma$} (s) \\ \toprule
        Decryption & 20 & 0.2865 & 0.2959 & 0.032 \\ \hline
        Encryption & 20 & 0.3155 & 0.3134 & 0.042 \\
        \bottomrule
    \end{tabular}
    \caption{Mobile benchmarking results}
    \label{tab:mobile}
\end{table}
\subsection{Server Benchmarking}
Server benchmarking proved to be a more difficult task than performance testing our mobile application. The biggest issue is that the server component uses asynchronous processes that are difficult to time. So instead, we found it best to utilize a network profiler, tcpdump, to gauge timing. Using tcpdump, we can determine the time from when the client first sends a new message for processing, to the point when the server returns the processed message back to the client. Since this timing also includes the period of time the mobile application is processing the message, we subtract the time the phone took from the overall timing. Doing this, we find the median, average ($\mu$), and standard deviation ($\sigma$) of times a message spent on the server in Table \ref{tab:server}.
\begin{table}[h]
    \centering
        \begin{tabular}{c | c c c c}
        \toprule
        & \textbf{$n$} & \textbf{median} (s) & \textbf{$\mu$} (s) & \textbf{$\sigma$} (s) \\ \toprule
        Decryption & 20 & 1.33 & 1.485 & 0.466 \\ \hline
        Encryption & 20 & 1.64 & 1.547 & 0.373 \\
        \bottomrule
    \end{tabular}
    \caption{Server benchmarking results}
    \label{tab:server}
\end{table}

\subsection{Client Benchmarking}
Performing a thorough benchmark on the client was not possible due to the lack of performance benchmarking tools for Mozilla Thunderbird. It was however, possible to perform benchmarks for small components of code that can in turn, give us a view of the total amount of time necessary to perform encryption actions with MEG. Namely, from Figure \ref{fig:Flow} we note that there are two parts where the client needs to interact with the server. These are: sending an AES encrypted message to the server and then polling the server for the processed message. In benchmarking both of these processes we can determine the amount of time, non-overlapping with the server, that it takes to perform actions on the Thunderbird client. To be as thorough as possible, we also benchmark the time used on AES encryption and decryption.
\begin{table}[h]
    \centering
        \begin{tabular}{c | c c c c c}
        \toprule
        & \textbf{Task} & \textbf{$n$} & \textbf{median} (s) & \textbf{$\mu$} (s) & \textbf{$\sigma$} (s) \\ \toprule
        Decryption & AES & 20 & 0.006 & 0.007 & 0.0037 \\ \hline
        & total & 20 & 0.168 & 0.182 & 0.0398 \\ \midrule
        Encryption & AES & 20 & 0.004 & 0.004 & 0.0003 \\ \hline
        & total & 20 & 0.088 & 0.209 & 0.1755 \\
        \bottomrule
    \end{tabular}
    \caption{Client benchmarking results}
    \label{tab:client}
\end{table}
\subsection{Aggregate Timing}
If we tally the total amount of time that occurred in all trials for both encryption and decryption of messages we can find the median, average ($\mu$), and standard deviation ($\sigma$) of the total time MEG used to process a message.
\begin{table}[h]
    \centering
        \begin{tabular}{c | c c c c}
        \toprule
        & \textbf{$n$} & \textbf{median} (s) & \textbf{$\mu$} (s) & \textbf{$\sigma$} (s) \\ \toprule
        Decryption & 20 & 1.808 & 1.962 & 0.462 \\ \hline
        Encryption & 20 & 2.098 & 2.07 & 0.374 \\
        \bottomrule
    \end{tabular}
    \caption{Aggregated benchmarking results}
    \label{tab:aggregate}
\end{table}
As we can see from Table \ref{tab:aggregate}, on average, any kind of message processing takes about 2 seconds to perform. We do not believe that users will be frustrated by this timing, however, usability studies are necessary to definitively prove this point.
\section{Future Work}
\par We have described how MEG is a secure and accessible application for encrypting emails. Future research efforts should focus on making the software as usable as possible. The requirement to scan a QR code may be onerous for some users; and there may be other routes to transfer symmetric key information, so this question is worth investigation. It has also been suggested that IPv6 would enable us to eliminate the server altogether as a message router allowing the client plugin to directly communicate with the phone. This also deserves further investigation.
\par Currently, MEG is in a proof-of-concept stage and we hope to make it accessible to as many people as possible in the future. To improve on accessibility we wish to conduct usability studies to alleviate any UI and user experience (UX) issues in MEG. One of the biggest remaining items is to create a colorized system for PGP \textit{web of trust} to allow users to quickly validate their communications and to reduce any lingering confusion about the purpose of \textit{web of trust}. The performance analysis shows that the slowest component of our system is the server. Reducing the number network calls and performing caching of commonly used, non-sensitive items like PGP public keys could help reduce this time. Additional work will consist of hardening the system against errors that may occur from network instability or user error. We also hope to add an iOS app and a Gmail client to widen MEG's potential user base.

\section{Conclusion}
\par MEG is a free, email client agnostic, encryption application that automates all the manual steps of PGP email encryption. MEG ensures that users do not have to worry about the details of securing their messages, making it both more secure and more accessible than other email encryption schemes. MEG developers have the benefit of learning from its predecessor applications. We have the ability to understand where previous schemes such as KCM failed. We also have the benefit of being able to use the ubiquity of smartphones to serve as email gateways. Network effects that prevent ubiquitous encryption can be resolved by ensuring that MEG is as usable as possible. We believe that users should not have to choose between security and usability. It is, for the average user, a matter of engineering the correct solution. We believe MEG is the best solution for email encryption.

\bibliographystyle{plain}
\bibliography{main.bbl}

\begin{thebibliography}{10}

\bibitem{altun2006clustering}
Oguz Altun, Nilgun Dursunoglu, and Mehmet~Fatih Amasyali.
\newblock Clustering application benchmark.
\newblock In {\em IISWC}, pages 178--181, 2006.

\bibitem{callegati2009man}
Franco Callegati, Walter Cerroni, and Marco Ramilli.
\newblock {Man-in-the-Middle Attack to the HTTPS Protocol}.
\newblock {\em IEEE Security and Privacy}, 7(1):78--81, 2009.

\bibitem{ciphermail-gateway}
ciphermail.
\newblock Email encryption gateway.
\newblock \url{https://www.ciphermail.com/gateway.html}.
\newblock Accessed: 2016-02-17.

\bibitem{qrcode-key-distribution}
John Denker.
\newblock {Distributing PGP Keys and Fingerprints}.
\newblock \url{https://www.av8n.com/computer/htm/distributing-keys.htm}.
\newblock Accessed 2016-05-21.

\bibitem{dingledine2006anonymity}
Roger Dingledine and Nick Mathewson.
\newblock {Anonymity Loves Company: Usability and the Network Effect.}
\newblock In {\em WEIS}, 2006.

\bibitem{downs2006decision}
Julie~S Downs, Mandy~B Holbrook, and Lorrie~Faith Cranor.
\newblock Decision strategies and susceptibility to phishing.
\newblock In {\em Proceedings of the second symposium on Usable privacy and
  security}, pages 79--90. ACM, 2006.

\bibitem{wired-encrypted-traffic}
Klint Finley.
\newblock {Encrypted Web Traffic More Than Doubles After NSA Revelations}.
\newblock \url{http://www.wired.com/2014/05/sandvine-report/}.
\newblock Accessed: 2016-05-20.

\bibitem{eff-scorecard}
Electronic~Frontier Foundation.
\newblock {Secure Messaging Scorecard}.
\newblock \url{https://www.eff.org/secure-messaging-scorecard}.
\newblock Accessed: 2016-02-17.

\bibitem{furnell2013usable}
Steven~M Furnell, Nathan Clarke, Cristian Thiago~Moecke, and Melanie Volkamer.
\newblock Usable secure email communications: criteria and evaluation of
  existing approaches.
\newblock {\em Information Management \& Computer Security}, 21(1):41--52,
  2013.

\bibitem{garfinkel2005make}
Simson~L Garfinkel, David Margrave, Jeffrey~I Schiller, Erik Nordlander, and
  Robert~C Miller.
\newblock {How to make secure email easier to use}.
\newblock In {\em Proceedings of the SIGCHI conference on human factors in
  computing systems}, pages 701--710. ACM, 2005.

\bibitem{garfinkel2005johnny}
Simson~L Garfinkel and Robert~C Miller.
\newblock {Johnny 2: a user test of key continuity management with S/MIME and
  Outlook Express}.
\newblock In {\em Proceedings of the 2005 symposium on Usable privacy and
  security}, pages 13--24. ACM, 2005.

\bibitem{gmail-tls-report}
Google.
\newblock {Email encryption in transit}.
\newblock \url{https://www.google.com/transparencyreport/saferemail/}.
\newblock Accessed: 2016-05-20.

\bibitem{gregg2013systems}
Brendan Gregg.
\newblock {\em Systems Performance: Enterprise and the Cloud}.
\newblock Pearson Education, 2013.

\bibitem{enigmail-handbook}
Ludwig Hugelschafer, Daniel Raffo, Patrick Brunschwig, and Robert~J. Hansen.
\newblock {OpenPGP Email Security for Mozilla Applications; The Handbook V
  1.8}.
\newblock
  \url{https://www.enigmail.net/documentation/Enigmail_Handbook_1.8_en.pdf}.
\newblock Accessed 2016-05-23.

\bibitem{hushmail}
Hushmail.
\newblock {How Hushmail Works}.
\newblock \url{https://www.hushmail.com/about/technology/how-it-works/}.
\newblock Accessed: 2016-02-17.

\bibitem{qrcode-authentication}
Yung-Wei Kao, Guo-Heng Luo, Hsien-Tang Lin, Yu-Kai Huang, and Shyan-Ming Yuan.
\newblock {Physical access control based on QR code}.
\newblock In {\em Cyber-enabled distributed computing and knowledge discovery
  (CyberC), 2011 International Conference on}, pages 285--288. IEEE, 2011.

\bibitem{force-use-of-self}
Howard Lightstone.
\newblock {How can I force use of self-signed SSL certificate}.
\newblock
  \url{https://productforums.google.com/forum/#!topic/gmail/6gODk9n65ZU}.
\newblock Accessed: 2016-02-17.

\bibitem{openkeychain}
OpenKeychain.
\newblock About.
\newblock \url{https://www.openkeychain.org/about/}.
\newblock Accessed: 2016-05-19.

\bibitem{sheng2006johnny}
Steve Sheng, Levi Broderick, Colleen~Alison Koranda, and Jeremy~J Hyland.
\newblock {Why johnny still can’t encrypt: evaluating the usability of email
  encryption software}.
\newblock In {\em Symposium On Usable Privacy and Security}, pages 3--4, 2006.

\bibitem{trusov2009effects}
Michael Trusov, Randolph~E Bucklin, and Koen Pauwels.
\newblock Effects of word-of-mouth versus traditional marketing: findings from
  an internet social networking site.
\newblock {\em Journal of marketing}, 73(5):90--102, 2009.

\bibitem{whitten1999johnny}
Alma Whitten and J~Doug Tygar.
\newblock {Why Johnny Can't Encrypt: A Usability Evaluation of PGP 5.0.}
\newblock In {\em Usenix Security}, volume 1999, 1999.

\bibitem{zimmermann1995official}
Philip~R Zimmermann.
\newblock {\em {The official PGP user's guide}}.
\newblock MIT press, 1995.

\end{thebibliography}
\end{document}